# Multiparty Probability Computation and Verification

Subhash Kak

**Abstract:** A multiparty computation protocol is described in which the parties can generate different probability events that is based on the sharing of a single anonymized random number, and also perform oblivious transfer. A method to verify the correctness of the procedure, without revealing the random numbers used by the parties, is proposed.

1. **Introduction**

Imagine a group of individuals who wish to generate a probability event that is to be mapped to specific payoffs [1],[2]. The motivation for this could be the casino or a variety of situations in multiparty computing [3]-[5]. If the event is the specific value of a random variable, how are the individuals to be certain that the procedure is honest? Such events can be generated by random physical processes that are either classical [6]-[11] or quantum [12]-[16]. But since physical processes are associated with a specific location they are often not convenient for applications where the individuals may be computer nodes at distant places. Bu even a physical process must have performance that can be mathematically established for it will interface with other systems that can be attacked. In practice, we seek decentralized or hybrid procedures which can be shown to be mathematically secure.

In some cases, two parties may mutually arrive at a probability event using the decentralized oblivious transfer (OT) protocol which is based on hardness of certain number-theoretic mappings [17],[18]. In all such cases it must be assumed that the two parties are authenticated to each other in order to forestall man-in-the-middle (MIM) and other attacks [3]-[5],[19],[20]. In order to detect cheaters, one needs a verification procedure that does not reveal information on the random number generators used by the communicating parties for that could compromise the security of the system as was shown recently for a two-party situation [19]. Verification could also be based on secret sharing systems [21]-[23] but that constitutes a different kind of a system that will not be investigated here.

In this paper we first present a protocol for sharing a random number between more than two parties. Then we present a general verification procedure for multiparty probability events that involves oblivious transfer. The procedure consists of three parts: first, the set-up where initial information is exchanged by the parties to agree on a random number; second, a mapping process that takes the input to the range [0,1] with uniform probability; and third, the verification process.

2. **Initial set up**

Consider communicating parties Alice, Bob, and Charlie (the list can be augmented but here for simplicity we only speak of three) who wish to perform a secure computation, which is the sharing of random number. The first thing to be done is to create aliases so that actions within the computation are protected by the complexity of the computation. Each of these aliases is a random number. The three also wish to generate a single number that connects them with the multiparty computation.

In a centralized system (Figure 1), the trusted authority T performs the computation on the numbers *a, b, c* sent respectively by Alice, Bob, and Charlie. The numbers should be sent to T in a manner that hides each sender's identity. This requires a privacy preserving transformation where this hiding is accomplished by means of an appropriate one-way function.

Let the transformation carried out by T map the numbers to the range, R, which is [0,1]:

$$R = T(a,b,c) \qquad (1)$$

R maps to different probabilities $p_{ALICE}, p_{BOB}, p_{CHARLIE}$ for the three communicating parties. This mapping may be done by assigning non-overlapping one-thirds of the range [0,1] to the three parties.

$$p_{ALICE}, p_{BOB}, p_{CHARLIE} = f_i(R) \qquad (2)$$

The difficulty with this centralized procedure is that the users do not know if the transformation T is good at randomization. Although there is no way for them to confirm that the output R has a distribution which is uniform over [0,1], a strong hashing function will be considered satisfactory in most cases. Centralized procedures are implemented in many computer-controlled applications like the ones in a casino or in online gambling. In these latter applications, the assignment of probabilities is determined by the nature of the computation (or game) and the house is also assigned a certain portion of the take in accordance with law.

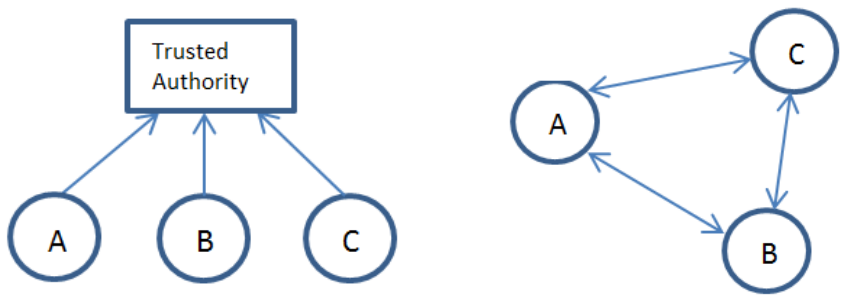

Figure 1. Centralized system with trusted authority; (right) decentralized system

The difference between the centralized and the decentralized systems is shown in Figure 1. In a decentralized system, after the users have been authenticated by some other protocol, they will send their random numbers *a, b,* and *c* to each other. The DH protocol [24], based on the use of a large prime *p* and integer *u* of large order modulo *p,* may be used by the three to exchange numbers between each pair. This procedure is more than just a pairwise exchange of random numbers as in the standard DH protocol, since a product of the three must also be exchanged. This latter step is required before functions like that of (1) and (2) may be implemented.

**The Protocol**

The protocol begins with a pairwise exchange of random numbers and then the product of the three:

*Step 1.* Alice and Bob share $u^{ab} \bmod p$, Bob and Charlie share $u^{bc} \bmod p$, and Charlie and Alice share $u^{ac} \bmod p$. (Figure 2)

*Step 2.* Bob sends $u^{ab} \bmod p$ to Charlie, who sends $u^{bc} \bmod p$ to Alice, who sends $u^{ac} \bmod p$ to Bob.

*Step 3.* Using their secret numbers, each is now able to compute the same key to be shared amongst them which is $u^{abc} \bmod p$. (Figure 3)

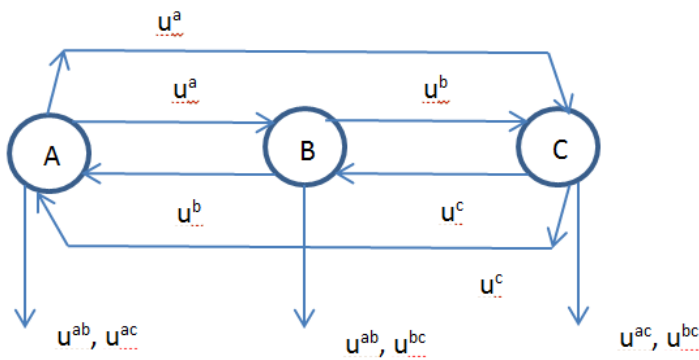

Figure 2. Pairwise exchange of random numbers

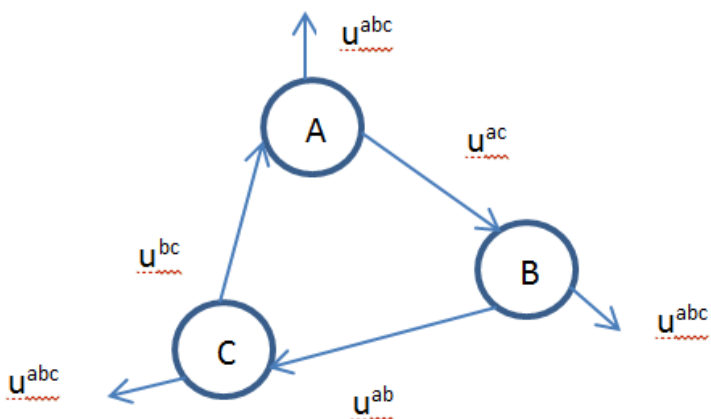

Figure 3. Generation of the single key $u^{abc} \bmod p$

As is clear from the working of this protocol as shown in Figures 2 and 3, the pairwise sharing of numbers as well as the final generation of a single number can be generalized to any number of parties. As is to be expected, the results do depend on the order in which the updating is done.

If the sharing of the random number is for the award of a prize, the participants are assigned random numbers and a standard hashing application. For example, for three participants, the parties randomly called 1, 2, and 3, will compute the hash of $u^{abc}$ and sum the two least significant bits of the hash. To account for the discrepancy between four cases and only three players, one may use the mapping:

> 00 – repeat the algorithm
> 01 – Player 1 wins
> 10—Player 2 wins
> 11—Player 3 wins

In the case of $2^k$ players, if a similar mapping was used, one would not need to repeat the process.

If one wished to use this protocol to generate oblivious transfer then the parties should randomly choose between a set of potential bases as in Step 4.

> *Step 4.* The three parties choose from different public numbers of larger order mod *p*. We will call these *u, v, w, ….*

We can imagine that the payoff is a prize that will be locked by the key generated by Alice and, therefore, available to those who get to share the key. In other words, this case can have multiple winners.

*Example 1.* Consider that the base integers used by the three are two in number and let's call them *u* and *w* (Figure 4). We will show that the two parties who pick the same base do not end up with the shared secret between them.

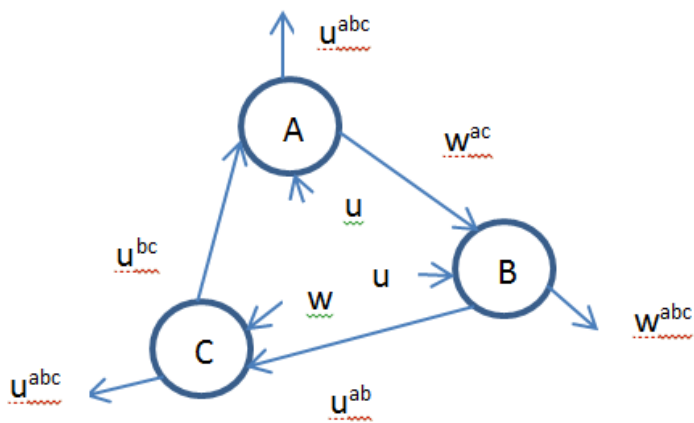

Figure 4. Generation of key when there are two bases

The rows of Table 1 going from top to bottom show intermediate steps in the formation of the key. In this example, the user whose pick is in the minority ends up getting the shared key.

Table 1.

| A | B | C |
|---|---|---|
| u | u | w |
| $w^{ac}$ | $u^{ab}$ | $u^{ac}$ |
| $u^{abc}$ | $w^{abc}$ | $u^{abc}$ |

None of the parties will know which one is the odd one out in the choice of the base. Alice does not know whether Bob or Charlie have the shared key, or if both of them have it (which is the case if they all pick the same base). Summarizing, we can sum it up in Table 2 (where the first three columns show the initial choice of the bases):

Table 2.

| A | B | C | Result |
|---|---|---|---|
| u | u | u | A, B and C share key |
| u | u | w | A, C share key |
| u | w | u | B, C don't share key with A |
| w | u | u | A, B share key |

In the above 4 cases, A shares key with one of the other two (or both) in three out of four cases. These could be compared to the four sequences 11, 01, 00, 10, where 1 represents sharing and 0 represents not sharing.

3. **The case of four parties**

In the case of four parties and two bases (*u* and *w*), the following cases will be different:

i. All chosen bases are the same (in which case the keys would be identical)
ii. Three choose one base and the fourth chooses another
iii. Two adjacent parties choose one base and the other two pick a different one
iv. Two non-adjacent parties choose one base and the other two pick the other

The cases ii, iii, and iv are described by Tables 3, 4, and 5, respectively.

Table 3.

| A | B | C | D |
|---|---|---|---|
| u | u | u | w |
| $w^{ad}$ | $u^{ab}$ | $u^{bc}$ | $u^{cd}$ |
| $u^{acd}$ | $w^{abd}$ | $u^{abc}$ | $u^{bcd}$ |
| $u^{abcd}$ | $u^{abcd}$ | $w^{abcd}$ | $u^{abcd}$ |

Table 4.

| A | B | C | D |
|---|---|---|---|
| u | u | w | w |
| $w^{ad}$ | $u^{ab}$ | $u^{bc}$ | $w^{cd}$ |
| $w^{acd}$ | $w^{abd}$ | $u^{abc}$ | $u^{bcd}$ |
| $u^{abcd}$ | $w^{abcd}$ | $w^{abcd}$ | $u^{abcd}$ |

Table 5.

| A | B | C | D |
|---|---|---|---|
| u | w | u | w |
| $w^{ad}$ | $u^{ab}$ | $w^{bc}$ | $u^{cd}$ |
| $u^{acd}$ | $w^{abd}$ | $u^{abc}$ | $w^{bcd}$ |
| $w^{abcd}$ | $u^{abcd}$ | $w^{abcd}$ | $u^{abcd}$ |

In case (ii), B and D share the key with A; in case (iii), only D shares the key with A; and in case (iv), C shares the key with A. Since the key generation process has three steps (represented by the three bottom rows of each table), the base travels one step to the right at each stage, ending up 3 positions to the right which is equivalent to one position to the left.

In Table 3, the total favorable probability of one of the three (B,C,D) obtaining the same key as A is 4/9 as shown in Table 6:

Table 6.

| A | B | C | D | Result |
|---|---|---|---|---|
| u | u | u | w | A, B, and D share key |
| u | u | w | u | A, C, and D share key |
| u | w | u | u | B, C, and D don't share key with A |
| w | u | u | u | A, B, and C share key |

If sharing of key with A by B, C, and D is represented by 1, these four cases represent the sequences 101, 010, 000, and 110. The cases of Table 4 map to the sequences 001, 100, 011, and that of Table 5 to the sequence 010.

Clearly, such analysis can be extended to more general cases.

4. **Verification process for three base integers**

Now consider that there are three base integers, *u, v,* and *w*. To forestall cheating by any party, one would need to develop a verification sequence by using a previously announced random number r that is used as an exponent on the respective raw keys.

Consider the sequence $G(n) = u^n + v^n + w^n \mod p$. To relate the three variables amongst each other, we need a quadratic expansion of the kind below:

$$u^3 = \alpha u^2 + \beta u + \gamma \mod p$$
$$v^3 = \alpha v^2 + \beta v + \gamma \mod p$$
$$w^3 = \alpha w^2 + \beta w + \gamma \mod p \qquad (3)$$

This may be written down as the matrix equation:

$$\begin{bmatrix} u^3 \\ v^3 \\ w^3 \end{bmatrix} = \begin{bmatrix} u^2 & u & 1 \\ v^2 & v & 1 \\ w^2 & w & 1 \end{bmatrix} \begin{bmatrix} \alpha \\ \beta \\ \lambda \end{bmatrix}$$

The solution of this equation is easily found to be:

$$\begin{bmatrix} \alpha \\ \beta \\ \gamma \end{bmatrix} = \frac{1}{(v-w)(w-u)(u-v)} \begin{bmatrix} v-w & w^2-v^2 & v^2w-vw^2 \\ w-u & u^2-w^2 & uw^2-wu^2 \\ u-v & v^2-u^2 & u^2v-uv^2 \end{bmatrix}^T \begin{bmatrix} u^3 \\ v^3 \\ w^3 \end{bmatrix} \quad (4)$$

This may be simplified to

$$\alpha = (u+v+w); \beta = -(uv+vw+wu); \gamma = uvw \quad (5)$$

**Theorem 1**. $G(n) = \alpha G(n-1) + \beta G(n-2) + \gamma G(n-3) \bmod p$ \quad (6)

*Proof.* $G(n) = (u^n + v^n + w^n) \bmod p$
$= (u^{n-3}u^3 + v^{n-3}v^3 + w^{n-3}w^3) \bmod p$
$= u^{n-3}(\alpha u^2 + \beta u + \gamma) + v^{n-3}(\alpha^2 v + \beta v + \gamma) + w^{n-3}(\alpha^2 w + \beta w + \gamma) \bmod p$
$= \alpha G(n-1) + \beta G(n-2) + \gamma G(n-3) \bmod p$

The sum of successive powers of *v* and *w* suffices to establish that they have been computed to the same exponent. All that is required to find the values of α and β is the solution to equation (3) for *k* = 2. No knowledge of the actual value of n is needed while computing equation (6).

*Example 2.* Let *u*=2, *v*=3, and *w*=5 mod 17. To find α, β, and γ, we use equation (5), obtaining:

$\alpha = 10; \beta = 3; \gamma = 13 \bmod 17$

The series $G(n) = 2^n + 3^n + 5^n \bmod 17$, for n = 0, 1, 2, 3,… is as follows:

3, 10, 4, 7, 8, 0, 13, …

for which each *n*th element is 10 G(n-1)+3G(n-2) +13G(n-3) mod 17. For example, the value 13 is 10×0+3×8+13×7 mod 17.

5. **Conclusions**

This paper considers the problem of generation of random events. A multiparty computation protocol is described in which the parties can generate different probability events, which is achieved by sharing a single anonymized random number. The paper also describes a method of multiparty oblivious transfer using DH protocol. Several specific cases of two bases used by three or four parties are considered and the cases where the parties will end up sharing the key with the generating party are identified.

A method to verify the correctness of the procedure, without revealing the random numbers used by the parties, is also proposed.

*Acknowledgement*. This research was supported in part by research grant #1117068 from the National Science Foundation.